\documentclass[twocolumn, aps]{revtex4}

\usepackage{epsfig}

\begin{document}

\title{Hydrophobic Interaction Model for Upper and Lower Critical Solution Temperatures}
\author{Susanne Moelbert$^1$}
\email{susanne.moelbert@ipt.unil.ch}
\author{Paolo De Los Rios$^{1,2}$}
\affiliation{$^1$Institut de Physique Th\'eorique, 
Universit\'e de Lausanne, CH-1015 Lausanne, Switzerland.\\
$^2$INFM Sezione di Torino - Politecnico, 
Corso Duca degli Abruzzi 24, 10129 Torino, Italy.}
\date{\today}

\begin{abstract}
Hydration of hydrophobic solutes in water is the cause of different phenomena, including the hydrophobic heat-capacity anomaly, which are not yet fully understood. Because of its topicality, there has recently been growing interest in the mechanism of hydrophobic aggregation, and in the physics on which it is based. In this study we use a simple yet powerful mixture model for water, an adapted two-state Muller-Lee-Graziano model, to describe the energy levels of water molecules as a function of their proximity to non-polar solute molecules. The model is shown to provide an appropriate description of many-body interactions between the hydrophobic solute particles. The solubility and aggregation of hydrophobic substances is studied by evaluating detailed Monte Carlo simulations in the vicinity of the first-order aggregation phase transition. A closed-loop coexistence curve is found, which is consistent with a mean-field calculation carried out for the same system. In addition, the destabilizing effect of a chaotropic substance in the solution is studied by suitable modification of the MLG model. These findings suggest that a simple model for the hydrophobic interaction may contain the primary physical processes involved in hydrophobic aggregation and in the chaotropic effect. 
\end{abstract}
\maketitle

\section{Introduction}
Many recent studies have focused on the surprising thermodynamic properties associated with the hydration of hydrophobic substances in aqueous solutions~\cite{tanford, pollack}. Indeed, the heat capacity of non-polar solutes in water is large and positive at room temperature. This behavior, known as the ``hydrophobic heat-capacity anomaly'', stands in sharp contrast to the observations made for hydrophilic solutes in water~\cite{ben-naim1}. Experiments confirm that the solubility of small hydrocarbons in water decreases when the solution is heated near room temperature. It is generally believed that these thermodynamic phenomena are associated with a structural change in the solvent, and the hydrophobic effect is therefore considered as directly related to the anomalous properties of liquid water~\cite{ben-naim2}. 

Due to hydrogen bonds between molecules, pure water is highly structured. Adding a non-polar solute, which is unable to form hydrogen bonds, disrupts the structural arrangement by breaking bonds, thus raising the enthalpy of the solution and increasing considerably the entropy. Because the entropy change dominates over the enthalpy change at room temperature, the hydration process is considered to be ``entropy driven''. 

Although this intuitive interpretation provides a first understanding of the hydrophobic effect, it still suffers from a number of limitations. In contrast to the expected increase in enthalpy due to the breaking of hydrogen bonds, the process of transferring a non-polar particle to water from the liquid state in pure solute is found experimentally to be enthalpically even slightly favorable (negative), at least at low temperatures~\cite{privalov}. This phenomenon can be explained only by a rearrangement of the water molecules in the vicinity of the solute, resulting in a recovery of the lost hydrogen bonds which tend to be slightly stronger than before~\cite{frank}. This new structure, however, causes an entropy loss attributed to an increase in local order. Measurements of the entropy and enthalpy changes for different solute sizes suggest that this local ordering of water molecules around the hydrophobic particle is not unique, but that a number of different organizations is possible~\cite{tanford}. Hence, the anomalous thermodynamic properties are a consequence of changes in the state of water molecules due to the insertion of hydrophobic solute particles, rather than being accounted for by water-solute interactions. 

At room temperature the solubility of small, hydrophobic solute species in water increases when the system is cooled. At sufficiently low temperatures a homogeneous mixture is found, provided that none of the components crystallizes. Heating the mixture produces a decrease in solubility, which can result in a phase transition to a state with two phases of different solute densities. This transition temperature is called the Lower Critical Solution Temperature (LCST), and has been measured for different solutions of hydrophobic particles in water~\cite{davies, bae, schild, costa}. After reaching a minimum, the solubility increases steeply at higher temperatures. If a coexistence of two phases is found near room temperature, another phase transition to one homogeneous phase occurs at the Upper Critical Solution Temperature (UCST), on condition that the boiling point of the solution is not reached~\cite{yalkowsky}. 

Aqueous solutions of non-polar particles which show a LCST also have an UCST and therefore a closed-loop coexistence curve in the phase diagram, provided that none of the liquids undergoes a phase transition to a gaseous phase before that temperature is reached. This occurs because the rising temperature increases the entropy, which favors a homogeneous mixing of the components. In addition, the number of hydrogen bonds formed, which is responsible for the phase separation, decreases with increasing temperature. This type of closed-loop miscibility curve has been found in different binary solutions including nicotine/water and poly(ethylene glycol)/water~\cite{davies, bae}, as well as by modeling of binary mixtures containing hydrogen-bonded molecules~\cite{barker, wheeler, walker}. 

A complete understanding of the molecular mechanisms underlying the hydrophobic effect is essential in order to explain a variety of phenomena, including the aggregation of hydrophobic solutes in water and the destabilization, denaturation and aggregation of proteins, processes which are responsible for many life-sustaining processes but also for different diseases~\cite{dill, wiggins, carrell}. 

Aggregation of non-polar particles in aqueous solutions can be attributed to the effective hydrophobic interaction between solute particles. This interaction is mainly solvent-induced, because the approach of two hydrophobic solute particles reduces the surface exposed to water compared with the total surface when they are completely separated. However, despite its importance for the understanding of a number of still-unexplained phenomena, the study of hydrophobic aggregation is at present incomplete and needs further investigation. In this work we describe and analyze solutions of hydrophobic particles, particularly in the vicinity of the aggregation phase transition, using a simple two-state model in three dimensions. The results of a mean-field calculation are compared to Monte Carlo simulations. In addition, the model is extended to include the effect of a chaotropic substance in the solution, which causes an increase in solubility of the non-polar solute for reasons which are at present not well understood.

This paper is organized as follows. Section II introduces the model used for a detailed description of the solution. In Section III a mean-field calculation is performed, giving a first idea of the phase diagram near the aggregation transition. Monte Carlo simulations confirming these findings are presented in Section IV. In Section V the model is adapted to describe the chaotropic effect. Finally, Section VI offers a discussion of the results, as well as some suggestions for further investigations.

\section{Model}
\label{secModel}
In order to describe the interaction between hydrophobic solutes and water, a simple model which describes the essential physics of the system is required. It is generally believed that the driving force in the aggregation process is the effective repulsive hydrophobic interaction between the polar water and non-polar solute~\cite{kauzmann}. As early as 1945, Frank and Evans~\cite{frank} noted that this interaction arises from a cage-like arrangement of water molecules around the solute, which allows them to optimize their mutual hydrogen bonding and thus to minimize their energy with respect to bulk liquid water. Entropically, however, this strict ordering is unfavorable compared with the disordered configurations in bulk water.

The minimal model which contains the essential features of the physics of water as an aqueous solvent is the bimodal description of Muller, Lee, and Graziano (MLG)~\cite{muller, lee}. The model describes liquid water by dividing it into two different populations based on the number of hydrogen bonds formed. Water molecules which are highly hydrogen-bonded to their neighbors have fewer rotational degrees of freedom and thus a lower multiplicity of degenerate configurations (lower entropy) than unbonded molecules. These are therefore denoted as ``ordered'', while water molecules with many broken hydrogen bonds are considered to be ``disordered''. The presence of a non-polar solute alters the enthalpy and entropy of water molecules in the solvation shell of each solute particle, and so a further distinction is required between ``shell'' and ``bulk'' water. More details of the physical processes underlying these considerations are provided below. This simple model has been justified recently from molecular simulations of water~\cite{silverstein1}.

In this form the MLG model reproduces correctly the ordering and strengthening of the hydrogen bonds in the first solvation shell of an added non-polar solute molecule at low temperatures, as well as the opposite behavior at high temperatures. It has been shown that the model provides an adequate description of the heat-capacity anomaly~\cite{silverstein2}, and it has been used to reproduce consistently the important properties of protein solutions, including warm and cold denaturation~\cite{delos, caldarelli}.

In this analysis we use an adapted version of the MLG model. An appropriate description of the solvent in the vicinty of hydrophobic solute particles which may be much larger than individual water molecules is obtained by allowing each site of a lattice representing the system to be occupied by a group of water molecules. The bimodal nature of the MLG model is preserved in this coarse-grained version by specifying only two types of water cluster at each site, where now an ``ordered'' site is characterized by having most of the hydrogen bonds among the molecules in the cluster intact, while a ``disordered'' site is understood to have a number (but by no means all) of these bonds broken. While this is the simplest possible approximation to the continuous distribution of intact or broken bonds within a site, we will demonstrate that it retains the capability of describing all the primary physical properties observed in aqueous solutions of non-polar solutes. The energy and degeneracy parameters for the coarse-grained model are determined by the same processes as in the bare MLG description outlined above, which we now discuss in more detail.

An approximation where a group of water molecules is considered as one entity is justified when the non-polar solute particle is relatively large compared with a single water molecule. In this case the formation of a complete cage around the solute particle is rather improbable because it must be formed rapidly in the presence of local thermal fluctuations, and may even be prevented sterically~\cite{nemethy}. Partial cages may therefore be formed in the vicinity of a solute particle, rather than one complete cage. In addition, formation of a hydrogen bond promotes the formation of further hydrogen bonds, which are stronger than before due to the change in charge distribution on forming the first bond~\cite{stillinger80}. This mutual reinforcement is known as ``cooperativity'' of hydrogen bonds, and leads to the formation of chains or clusters of hydrogen-bonded water molecules, whose extent depends on the size and the shape of the solute particles. 

Water molecules which participate in hydrogen-bonded clusters have a higher degree of order and fewer rotational degrees of freedom than those in regions with many unbonded molecules~\cite{nemethy}. The number of possible configurations of such an ``ordered'' cluster is thus significantly smaller than that of a ``disordered'' group of water molecules for both shell and bulk water. For steric reasons, fewer hydrogen-bonded water configurations are possible around a non-polar solute particle which is unable to form hydrogen bonds. These shell water molecules are forced into a tangential orientation~\cite{deJong}, whereas the molecules in bulk water may also form radially oriented hydrogen bonds with central water molecules replacing the solute~\cite{silverstein2}. The degeneracy, or total number of configurations, of a hydrogen-bonded cluster of shell water (``ordered shell'') is consequently smaller than that of a hydrogen-bonded cluster of bulk water (``ordered bulk'').

In contrast, fewer orientational configurations exist for unbonded water molecules in the bulk than next to a non-polar solute particle~\cite{silverstein2}. The geometrical reason for this result is that for a shell site no hydrogen bonds are possible in direction of the solute particle, unlike the bulk situation obtained on replacing the 
solute particle by water. All orientations in which water molecules form radial hydrogen bonds with the central water in the bulk case (contributing to the ordered bulk degeneracy) are therefore transformed into configurations with many broken hydrogen bonds when the central water is replaced by a non-polar solute particle. We take such sites to be ``disordered'' in the bimodal sense discussed above. The degeneracy of a group of water molecules with many broken hydrogen bonds is then higher in the shell (``disordered shell'') than in the bulk (``disordered bulk''). 

In summary, these considerations lead to a distribution of the total number of states $q$ according to the sequence $q_{ds} > q_{db} > q_{ob} > q_{os}$. Because many fewer configurations have intact than broken hydrogen bonds, the difference in degeneracy between ordered and disordered states is much higher than that between shell and bulk states for both types of site. In fact the difference in degeneracy between shell and bulk states depends primarily on the number of possible radial hydrogen bonds, which is much smaller than the total number. In much of what follows we employ the degeneracy factors $q_{ds} = 49,\ q_{db} = 40,\ q_{ob} = 10$, and $q_{os}= 1$. These relative values are chosen to be qualitatively representative of the degeneracies expected for a system of small solute particles in water, with both experimental and theoretical~\cite{silverstein2} justification based on the above properties of water. We note here that, while it is known that a number of ordered states exists even for the cage-like water structures coordinating dissolved solute particles, the absolute multiplicities of the $q$ factors contribute only an additive constant to the free energy and are irrelevant for the phenomena to be discussed below; thus we set the lowest degeneracy to $q_{os} = 1$.

\begin{figure}[h!]
\includegraphics[width=8cm]{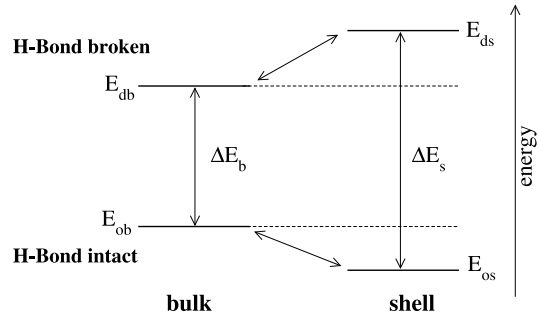}
\caption{Energy levels in the MLG model for water.  }
\label{FigMLG}
\end{figure}

Fig.~\ref{FigMLG} shows the schematic energy levels of a group of water molecules in the coarse-grained MLG-type model. The hydrogen-bond energy is optimized in the ordered, cage-like shell structure where strong, tangentially oriented hydrogen bonds are formed among water molecules in the cluster (``ordered shell''), and hence $E_{os}$ is lowest. Direct experimental evidence for the tangential orientation of hydrogen bonds between water molecules in the solvation shell of non-polar solute particles has been provided recently in Refs.~\cite{pertsemil, deJong}. When the solute particle is replaced by water, radial hydrogen bonds may be formed. However, clusters including radially oriented water molecules have on average a higher hydrogen-bond energy than those whose bonds are predominantly tangentially oriented, because for steric reasons a radial hydrogen bond precludes another good tangential hydrogen bond to a first-shell neighbor. This result was demonstrated in a model including oriented hydrogen bonds between water molecules~\cite{silverstein2}. In bulk water, both configurations are possible, and thus the average energy in a cluster of hydrogen-bonded bulk water $E_{ob}$ (``ordered bulk'') is higher than $E_{os}$. 

The energies $E_{ds}$ (``disordered shell'') and $E_{db}$ (``disordered bulk'') are relatively much higher than those of the respective ordered states, because breaking of hydrogen bonds is required in forming the disordered states. The average hydrogen-bond energy of a group of disordered shell water molecules decreases when the solute particle is replaced by water, because some radial hydrogen bonds broken by the solute may be formed, lowering the average energy of the group and ensuring that $E_{db} < E_{ds}$. 

Specific determination of the energy values for a selected binary system would require structural calculations and molecular dynamics simulations~\cite{delos}. However, such refinement is not necessary for the general phenomena to be illustrated below. We note here that the temperature scale $\beta^{-1}$ is defined by the energy scale. In the following we use the parameter values $E_{ds} = 1.8,\ E_{db}= 1.0,\ E_{ob}= -1.0$, and $E_{os} = -2.0$, which are thought to be quite generic for aqueous systems, and furthermore agree closely with the energies used in a successful description of the thermodynamic behavior of biopolymers in water~\cite{delos2}. While the results of the calculations to follow are not particularly sensitive to the exact values selected, it remains possible to refine these parameters by comparison with experiment to obtain good qualitative agreement with measured quantities for different solutions. 

On a cubic lattice the energy of a system of $N$ sites, occupied either by particles ($n_i=0$) or by water ($n_i=1$), is given by the Potts-like Hamiltonian
\begin{eqnarray}
\nonumber
H[\{n_i\},\{\sigma_i\}]& = &\sum_{i=1}^N n_i[(E_{os} \tilde\delta_{i, \sigma_{os}} + E_{ds}
\tilde\delta_{i, \sigma_{ds}})(1-\lambda_i) \nonumber \\
& &+ (E_{ob} \tilde\delta_{i, \sigma_{ob}} + E_{db} \tilde\delta_{i, \sigma_{db}}) \lambda_i],
\label{Hamiltonian}
\end{eqnarray}
where $\lambda_i$ is $1$ if site $i$ is surrounded only by water, is $0$ otherwise, and is defined as the product of the nearest-neighbor factors, $\lambda_i = \prod_{\langle j, i \rangle} n_j$. Each water site $i$ can be in one of the $q$ different states which are divided among the $4$ energy levels shown in Fig.~\ref{FigMLG}. Therefore, $\tilde\delta_{i, \sigma_{os}}$ is $1$ if site $i$ is occupied by water in one of the $q_{os}$ ordered shell states and $0$ otherwise, and $\tilde\delta_{i, \sigma_{ds}}$ is $1$ if it is occupied by water in one of the $q_{ds}$ disordered shell states and $0$ otherwise. The same holds for the bulk states.

An important observation is that this model does not include interactions between different water sites. Hence the model is valid as long as water is liquid, and neglects long-range effects arising from extended hydrogen-bonded networks.

Hamiltonian (\ref{Hamiltonian}) leads to the partition function
\begin{equation}
Z = \sum_{\{n_i\},\{\sigma_i\}} e^{-\beta H[\{n_i\},\{\sigma_i\}]},
\end{equation} 
where every term of the sum represents the statistical weight of the corresponding configuration ($\beta^{-1} = k_BT$) . Performing the sum over the configurations of the states of each water site, including the number of states of the respective energy levels, gives
\begin{equation} 
Z = \sum_{\{n_i\}} \prod_i e^{-\beta[{\cal B} n_i \lambda_i + {\cal S}n_i(1- \lambda_i)]},
\end{equation}
where we define
\begin{eqnarray}
{\cal S} &=& -\frac{1}{\beta} \ln[q_{os}e^{-\beta E_{os}} + q_{ds}
e^{-\beta E_{ds}}] \mbox{\ \ and} \\
{\cal B} &=& -\frac{1}{\beta}\ln[q_{ob}e^{-\beta E_{ob}} + q_{db} e^{-\beta E_{db}}].
\end{eqnarray} 
The formal method for obtaining an effective Hamiltonian (and effective interactions, see below) is to integrate over the degrees of freedom of the particles thought to be responsible for the interactions, which are the solvent molecules. The canonical partition function, {\it i.e.} the partition function for a fixed number of particles $N_p$ (whence the number of water sites is $N_w=N-N_p$), is
\begin{equation}
Z_{N_w} = \sum_{\{n_i\}} e^{-\beta H_{\rm eff}[\{n_i\}]},
\label{eq:ZN}
\end{equation} 
where the effective Hamiltonian $H_{\rm eff}[\{n_i\}]$ is formally the free energy of the solvent at fixed particle configuration. In a system where the number of particles is not fixed, a chemical potential $\mu$ associated with the replacement of water sites by solute particles ($\mu$ is thus the difference between the chemical potentials of water and solute particles) is included, and the grand canonical partition function may be expressed as
\begin{eqnarray}
\Xi &= &\sum_{N_w} e^{\beta \mu N_w} Z_{N_w}\\
 &= &\sum_{\{n_i\}} e^{-\beta H_{\rm eff}^{\rm gc}[\{n_i\}]}, \mbox{\ \ where} \label{eq:gcZ}\\ 
H_{\rm eff}^{\rm gc}[\{n_i\}] &= &\sum_{i=1}^N [ ({\cal S}+\mu) n_i + ({\cal B}-{\cal S}) n_i \lambda_i].
\label{eq:Heff}
\end{eqnarray}
$H_{\rm eff}^{\rm gc}[\{n_i\}]$ represents an effective Hamiltonian for single sites, and provides the first step in obtaining the effective interactions between particles. In order to describe the consequences for hydrophobic particles in the solution, we replace the number of water molecules $n_i$ by the number of particles $m_i$,
\begin{equation} 
n_i = 1-m_i,
\end{equation}
where $m_i$ is $1$ if the site is occupied by a particle and $0$ otherwise. With this substitution, $\lambda_i$ becomes the product over the nearest neighbors, $\lambda_i = \prod_{\langle j,i \rangle} (1-m_j)$, and takes the value $1$ only if site $i$ is completely surrounded by water molecules, or $0$ otherwise. Introducing an effective interaction $J_{\rm eff} = ({\cal B}-{\cal S})$ between particles, and the effective chemical potential $\mu_{\rm eff} = ({\cal S}+\mu)$, the effective Hamiltonian for the particles becomes
\begin{equation}
H_{\rm eff,\ p}^{\rm gc}[\{m_i\}] = K - \sum_{i=1}^N [ \mu_{\rm eff} m_i + J_{\rm eff} (m_i-1) \lambda_i],
\end{equation}
where $K=N({\cal S}+\mu)$. In this formulation it can be seen that the interactions are not limited to two-body terms, but include many-body interactions through the last term.

Because ${\cal S}$ is negative and decreases continuously, $\mu_{\rm eff}$ decreases as temperature increases, and the solubility drops. If $\mu>-{\cal S}$, $\mu_{\rm eff}$ is positive at low temperatures and the solubility is high. In the ground state ($T=0$), this results in a solid phase for $\mu>-{\cal S}$ and pure water for $\mu$ large and negative.

At low temperatures, $J_{\rm eff}$ is positive and therefore the interaction term is minimized by $\lambda_i=1$, which means that there is a repulsive force between particles. At high temperatures, however, increasing entropy effects cause $J_{\rm eff}$ to become negative, and the minimal interaction energy is obtained for $\lambda_i=0$, resulting in an attractive force between particles. These different effective forces result from the interplay of entropic and enthalpic effects, and give rise to the complex properties of solutions of hydrophobic particles.

\section{Mean-Field Calculation}
\label{secMF}
We model the system on a three-dimensional square lattice, where every site has $z=6$ nearest neighbors, which can be occupied either by a particle or by water. Our first approach is based on the assumption that spatial fluctuations of the density are insignificant. In order to make qualitative predictions, a mean-field approximation is used for the average occupancy of a site, $\langle n_i \rangle = \rho$, thus treating the density as constant throughout the system. (Because of the nature of the model, $\rho$ is connected to the number of water molecules, and therefore the solute particle density, which is used in the graphs to follow, becomes $\rho_p=1-\rho$.) The grand canonical mean-field free energy of a particle is given by
\begin{eqnarray} \nonumber
f &= &({\cal B}-{\cal S}) \rho^{z+1} + ({\cal S}+\mu)\rho \\
&+ &\beta^{-1} [\rho \ln \rho +(1-\rho) \ln(1-\rho)].
\label{eq:f}
\end{eqnarray}

\begin{figure}[h!]
\includegraphics[width=8cm]{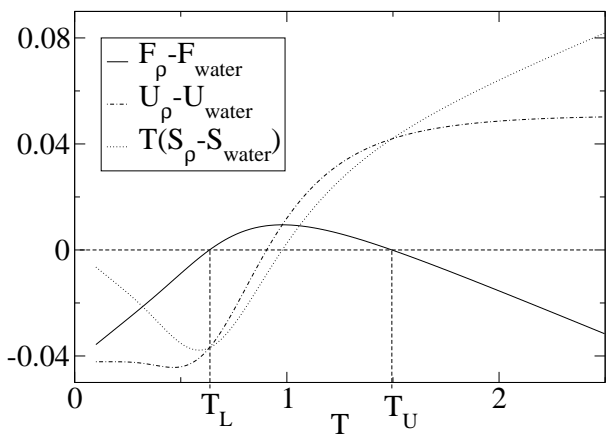}
\caption{Differences in free energy $F$, enthalpy $U$, and entropy $S$ (multiplied by $T$) per particle between a system of particle density $\rho_p = 0.1$ and pure water. The system with density $\rho$ is forced into a homogeneous state in which the particles are in solution. Between the critical temperatures $T_L$ and $T_U$, pure water is energetically favorable and the system separates into two phases.}
\label{FigDF}
\end{figure}

Comparison of the free energy per particle of a system with a given density to that of pure water (with the same number of water sites) illustrates the thermodynamic properties of the system. Because the density is fixed ($\mu = 0$), the system is forced into the homogeneous state although the free energy of a mixture of two separated phases with different densities might be lower. Pure water with the same number of water sites has approximately the same total separation of water and particles in a system of the same overall density. As shown in Fig.~\ref{FigDF}, a homogeneous mixture of density $\rho_p = 0.1$ is energetically favorable compared to pure water at low and at high temperature, indicating good solubility of the solute in water. In between, however, the solute prefers not to dissolve, and two phase transitions occur at the temperatures $T_L$ and $T_U$ where the differences in enthalpy and in entropy (multiplied by $T$) are equal. 

Because the system is forced into a homogeneous state, and $\mu=0$ in the calculation at constant $\rho_p$, the resulting free energy is larger than for a system minimizing $f(\rho)$ at a finite value of $\mu$ chosen to produce the same density $\rho_p$. In addition, the comparison with pure water neglects the small water-solute surface of a system in which the two species are completely separated, as it takes into account only the free energy of the water phase. The immiscible region therefore increases for a system allowed to separate into two phases. All of the following calculations are carried out by minimizing $f(\rho)$ in Eq.~(\ref{eq:f}), which provides the best available approximation.

\begin{figure}[h!]
\includegraphics[width=8cm]{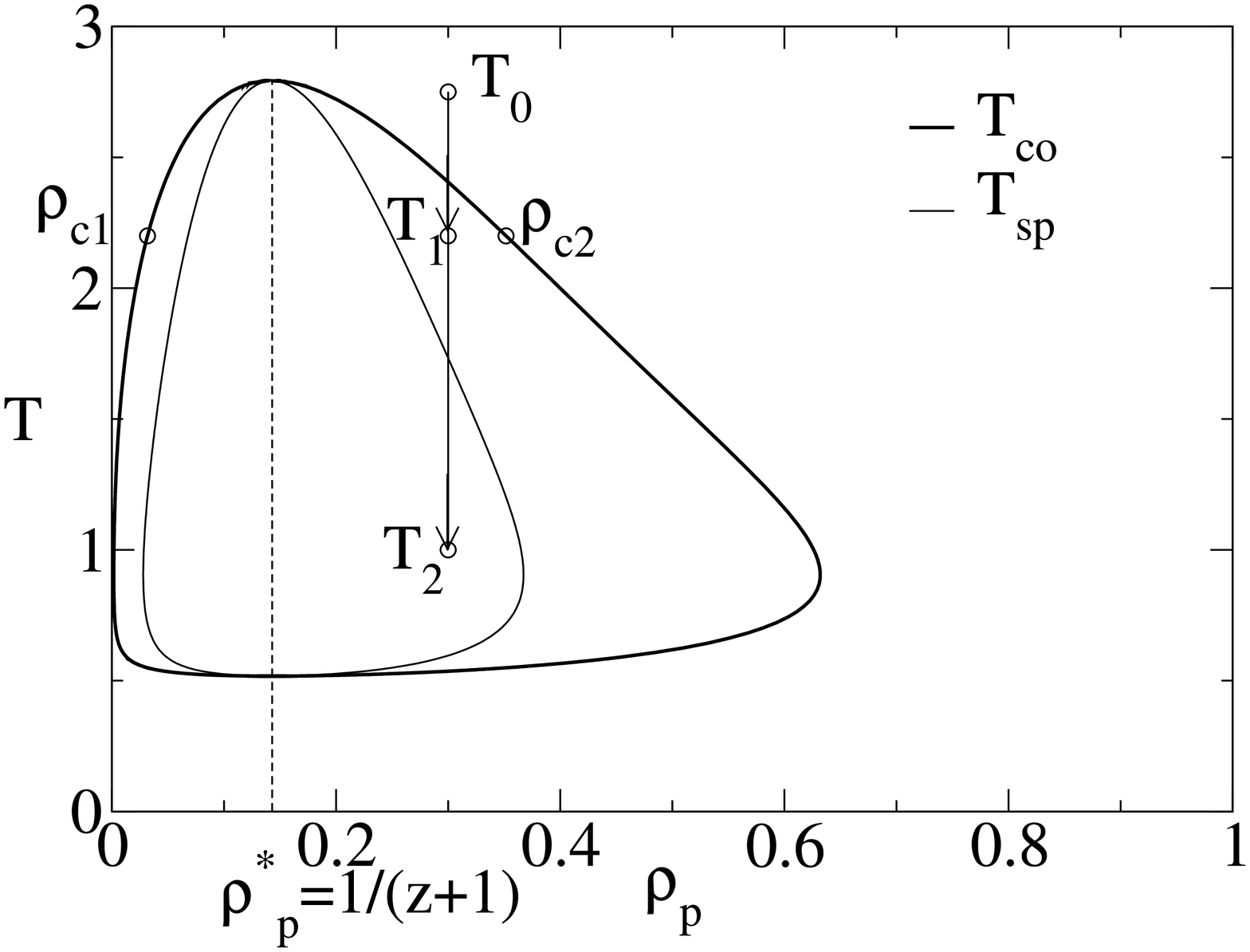}
\caption{$\rho$-$T$ phase diagram for an aqueous solution of non-polar solute particles obtained by mean-field calculation. The outer line represents the closed-loop coexistence curve, while the inner line marks the spinodal curve. The arrows of decreasing $T$ demarcate quenches into the metastable ($T_1$) and spinodal ($T_2$) regions (see text). }
\label{FigBouleMF}
\end{figure}

Based on these equilibrium values, a phase diagram for the values of $\rho$ can be obtained as a function of temperature, and is displayed in Fig.~\ref{FigBouleMF}. The outer line represents the closed-loop coexistence curve $T_{\rm co}(\rho)$, outside which the system is in a homogeneous state. Because we find aggregation by heating at low temperatures, and we expect that at high temperatures the entropy of solvation should dominate, a closed-loop solubility curve showing a LCST and an UCST~\cite{yalkowsky, bhimal, guillot} is to be anticipated. The inner line is the spinodal curve $T_{\rm sp}(\rho)$. 

Microscopically, the process of phase separation may be observed when quenching the system from the homogeneous phase at $T_0>T_U$, at a constant density $\rho_0$, into the coexistence region ($T_1$ or $T_2$). At long times after the quench, the system will be completely separated into two phases (with densities $\rho_{c1}$ and $\rho_{c2}$) in a ratio depending on the initial density $\rho_0$. The fraction of the total volume occupied by phase $i$ is $V_i = \frac{\rho_0 -\rho_j}{\rho_i-\rho_j}$, where $j$ stands for the other phase.

\begin{figure}[h!]
\includegraphics[width=8cm]{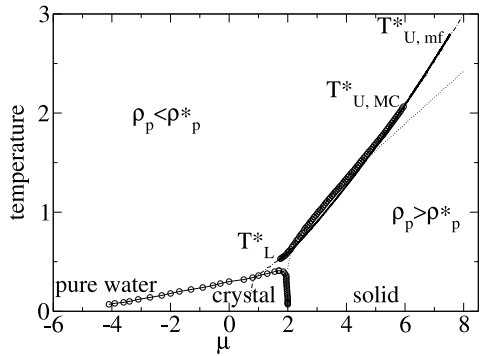}
\caption{Comparison of $\mu$-$T$ phase diagrams for hydrophobic particles in water obtained by mean-field calculation (solid line) and by Monte Carlo simulations ($\circ$). Both coexistence curves agree within the simulation error below the UCST of the Monte Carlo simulation $T_{\rm U, MC}^{\ast}$. The mean-field calculation results in a higher UCST ($T_{\rm U, mf}^{\ast}$) than do Monte Carlo simulations. A crystal phase appears at low temperatures in Monte Carlo simulations. The lines show a system heated at constant particle density $\rho_p=1/(z+1)$ ($-\!\cdot\!-$) and at $\rho_p=0.5$ ($\cdots$), corresponding to the vertical lines in Fig.~\ref{FigBouleMC}. Although the critical density, $\rho_p^{\ast}=1/(z+1)$, determined by mean-field calculation differs slightly from the value found by Monte Carlo simulations, it lies within the error, and no visible deviation is found from the expected behavior.  }
\label{FigMuT}
\end{figure}

Fig.~\ref{FigMuT} shows the phase diagram of the system a function of $\mu$ and $T$. On heating at constant $\mu$ (vertical lines) a phase transition is found at a temperature $T_L^* < T_t(\mu) < T_U^*$ (for $\mu_L^* < \mu < \mu _U^*$), where the particle density of the system jumps discontinuously from a value $\rho_{c2}$ to $\rho_{c1}$. At a constant temperature, such as $T_1$ in Fig.~\ref{FigBouleMF}, the free-energy density $f(\rho)$ is minimal at $\rho_{c1}$ and $\rho_{c2}$. Heating the system from below the transition temperature at fixed chemical potential $\mu$ results in a discontinuous jump in density from $\rho_{c1}$ to $\rho_{c2}$ at $T_t$.

The mean-field $\mu$-$T$ phase diagram in Fig.~\ref{FigMuT} shows a first-order transition line bounded by two critical points characterized by the same critical solvent density $\rho^{\ast}$. Analytically, $\rho^{\ast}$ may be obtained by imposing that the two local minima of the free energy $f$, as well as the inflection points of $f$, coincide at the critical points $\rho^{\ast}$, $T^{\ast}$ and $\mu^{\ast}$. The first, second and third derivatives of the free energy with respect to the density must therefore vanish at the critical points. We calculate simultaneously 
\begin{eqnarray}
\nonumber
\frac{\partial f}{\partial \rho} &= &(z+1)({\cal B}-{\cal S})\rho^z+({\cal S}+\mu)\\
& &+\beta^{-1} \ln(\frac{\rho}{1-\rho}) = 0,\label{eq:1stderiv}\\ \nonumber
\frac{\partial^2 f}{\partial^2 \rho} &= &z(z+1)({\cal B}-{\cal S})\rho^{z-1} \\
& & +\beta^{-1} \frac{1}{\rho(1-\rho)} = 0,  \label{eq:2ndderiv} \mbox{\ \ and}\\ \nonumber
\frac{\partial^3 f}{\partial^3 \rho} & =& z(z+1)(z-1)({\cal B}-{\cal S})\rho^{z-2}\\
	& &+\beta^{-1}( \frac{1}{(1-\rho)^2}-\frac{1}{\rho^2}) = 0,
\label{eq:3rdderiv}
\end{eqnarray}
and simplify Eq. (\ref{eq:2ndderiv}) to the form
\begin{equation}
\beta^{-1}=-z(z+1)(1-\rho)({\cal B}-{\cal S})\rho^z \label{eq:kT}.
\end{equation}
Introducing Eq.~(\ref{eq:kT}) into Eq.~(\ref{eq:3rdderiv}) leads to the critical density $\rho^{\ast} = \frac{z}{z+1}$ ({\it i.e.} critical particle density $\rho^{\ast}_p = \frac{1}{z+1}$), which depends only on the effective coordination number $z$. Inserting $\rho^{\ast}$ in Eq. (\ref{eq:kT}) provides the LCST $T_{L}^{\ast} = 0.518$ and the UCST $T_{U}^{\ast} = 2.79$ for the parameter values chosen. Finally, Eq. (\ref{eq:1stderiv}) gives the corresponding lower and upper critical chemical potentials $\mu_{L}^{\ast} = 1.69$ and $\mu_{U}^{\ast} = 7.53$. These values are shown in the $\mu$-$T$ phase diagram in Fig.~\ref{FigMuT}.

\section{Monte Carlo Simulation}
To analyze the behavior of the system in more detail, we study a three-dimensional system of $N=27\,000$ sites with a random initial distribution of particles and water molecules. Periodic boundary conditions are used to eliminate boundary effects. With regard to finite-size effects, we have found that our results are robust with respect to changes in the  system size. Because the system has a large number of degrees of freedom, a  representative sampling of the high-dimensional phase space is necessary to estimate  thermal averages in the equilibrium state. An appropriate technique which takes into account the effect of statistical fluctuations on the system is a Monte Carlo simulation.

In a system of fixed density ({\it i.e.} in the canonical ensemble), every possible configuration $\{n_i\}$ has the statistical weight
\begin{equation}
w_c(\{n_i\}) = \frac{e^{-\beta H_{\rm eff}[\{n_i\}]}}{Z_N},
\end{equation}
where the partition function $Z_N$ of a system of $N$ particles is given by Eq.~(\ref{eq:ZN}). At equilibrium the system must satisfy the detailed-balance condition 
\begin{equation}
w_c(\{n_i\}) P_{n\rightarrow n'} = w_c(\{n'_i\}) P_{n'\rightarrow n},
\end{equation}
where $P_{n\rightarrow n'}$ is the transition rate from configuration $\{n_i\}$ to a new one $\{n'_i\}$. The relative probability to produce configuration $\{n'_i\}$ from the previous one $\{n_i\}$ thus becomes the ratio of the two weights,
\begin{equation}
r = \frac{w_c(\{n'_i\})}{w_c(\{n_i\})} = e^{-\beta (H_{\rm eff}[\{n'_i\}] - H_{\rm eff}[\{n_i\}])}
\end{equation}
and depends only on the difference in free energy between them. This transition probability is used in the Metropolis algorithm to generate configurations from previous ones. Specifically, the new configuration $\{n'_i\}$ is accepted if $r>1$, or if $r<1$ but larger than a random number uniformly distributed in the interval $[0,1]$.

If the new configuration is completely different from the previous one, the acceptance probability is rather low. The method therefore sweeps randomly through the system considering configurations which differ from the previous one only by single-site exchanges of particles and water. After a number of thermalization sweeps, during which the system relaxes towards equilibrium and no observables are calculated, the system is taken to be in equilibrium with only thermodynamic fluctuations present. Thermodynamic quantities are estimated by averaging over the configurations which are kept during a subsequent number of steps which is sufficiently large that a considerable portion of the total phase space is sampled. The decorrelation time of successive configurations in equilibrium is found to be lower than $10$ Monte Carlo steps (one Monte Carlo step corresponds to the consideration of every site in the system once), both in the coexistence phase and in the homogeneous phase. Measurements are taken only every $50$ Monte Carlo steps over a period of $500\,000$ steps after $100\,000$ initial relaxation steps. This process is repeated for $10$ different random initial configurations and the observations are averaged over these independent simulations. In the crystal phase (below) the decorrelation time of consecutive configurations is longer because of the very low temperature, and measurements are averaged over a larger number of independent simulations. 

In order to find the equilibrium density for a fixed temperature and chemical potential, a grand canonical sampling ({\it i.e.}~the number of particles is not constant) of the phase space is performed. The procedure is the same as in the canonical case, except that the weight of a configuration $\{n_i\}$ in the grand canonical ensemble is
\begin{equation}
w_{\rm gc}(\{n_i\}) = \frac{e^{-\beta H_{\rm eff}^{\rm gc}[\{n_i\}]}}{\Xi},
\end{equation}
where the grand canonical partition function $\Xi$ is given by Eq.~(\ref{eq:gcZ}).
This leads to a relative transition probability for configuration $\{n'_i\}$ from a
previous one $\{n_i\}$ which depends on the difference in free energy of the two configurations
\begin{equation}
r = e^{-\beta (H_{\rm eff}^{\rm gc}[\{n'_i\}] - H_{\rm eff}^{\rm gc}[\{n_i\}])}.
\end{equation}

The behavior of a system which is heated at constant density can be analyzed by first determining the equilibrium density using grand canonical sampling at a chosen starting temperature and then continuously raising the temperature while applying a canonical sampling procedure in which the number of particles remains constant but particles and water may exchange sites.

Starting from a low temperature $T_0$ at fixed chemical potential $\mu$, we observe the thermal equilibrium state of the system at progressively higher temperatures:  for the same parameters as in Secs.~\ref{secModel} and \ref{secMF} a jump in the density from $\rho_{c1}$ to $\rho_{c2}$ occurs for values of $\mu$ between $\mu_L^* = 1.8$ (at the lower critical temperature $T_L^*=0.54$) and $\mu_U^* = 5.95$ (at the upper critical temperature $T_U^*=2.07$) (see Fig.~\ref{FigMuT}).

\begin{figure}[h!]
\includegraphics[width=8cm]{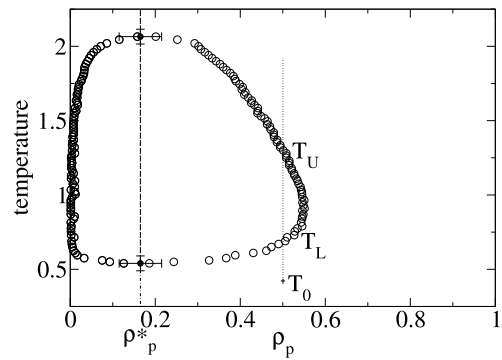}
\caption{$\rho$-$T$ phase diagram for an aqueous solution of hydrophobic particles obtained by Monte Carlo simulations, illustrating the coexistence curve. $\rho_p$ indicates the particle density. The system size is $N=27\,000$. The dotted line represents a system of density $\rho_p = 0.5$ which is heated from a starting temperature $T_0$. At $T_L$ a phase transition occurs from the homogeneous state to an aggregated phase. Further heating results in a second transition at $T_U$, where the system disaggregates and the particles are dissolved again.}
\label{FigBouleMC}
\end{figure}

Fig.~\ref{FigBouleMC} displays the $\rho$-$T$ phase diagram obtained from Monte Carlo simulations. The system shows a phase transition from a homogeneous state to a two-phase aggregation state at a lower transition temperature $T_L$, and a disaggregation at an upper transition temperature $T_U$, {\it i.e.} a closed-loop coexistence curve is found.

After rapid cooling at constant density $\rho_0$ from a temperature $T_0>T_U$ to a fixed temperature $T<T_U$ in the coexistence region, the system develops a clear phase separation into a phase with density $\rho_{c2}$ and nearly pure water. All of the particles aggregate to clusters, and after a certain period, one single cluster of density $\rho_{c2}$ remains, which occupies a fraction $V_{c2} = \frac{\rho_0 -\rho_{c1}}{\rho_{c2}-\rho_{c1}}$ of the total volume.

An analytical solution is possible for the ground state of the system at vanishing temperature, which provides a test for the Monte Carlo simulations. The calculation, which uses the fact that $E_{ds}>0$ and $E_{os}<0$, results in
\begin{eqnarray}
\nonumber
{\cal S}(T \rightarrow 0) &= &\lim_{\beta \rightarrow \infty}(-\frac{1}{\beta} \ln[q_{os}e^{-\beta E_{os}} +
q_{ds} e^{-\beta E_{ds}}])\\\nonumber
 &= &\lim_{\beta \rightarrow \infty}(-\frac{1}{\beta} [\ln(q_{os}) + \ln(e^{-\beta E_{os}}) ])\\
 &= &E_{os}
\end{eqnarray}
and, analogously, ${\cal B}(T \rightarrow 0)=E_{ob}$. At $T=0$, minimizing the free energy is equivalent to minimizing the Hamiltonian in Eq.~(\ref{eq:Heff}). This leads to three different phases depending on the chemical potential $\mu$. For $\mu>2.0$, the free energy is minimized by a solid phase ($\rho_p=1$ and $H=0$), while for $\mu<-5.0$ no particles are dissolved and pure water is found ($\rho_p=0$ and $H={\cal B}+\mu$). In between, the system forms a dispersed crystal structure in which the particles are arranged in such a way that every water molecule is the neighbor of exactly one particle, which leads to a particle density $\rho_p=\frac{1}{z+1}=1/7$ and $H=z({\cal S}+\mu)/(z+1)$. 

Monte Carlo simulations of the above system at sufficiently low temperatures confirm this behavior, as shown in the phase diagram in Fig.~\ref{FigMuT}. The dot-dashed line shows the evolution of a system which is heated at constant particle density $\rho_p=\frac{1}{z+1}$, which corresponds to the critical particle density $\rho_p^{\ast}$ found by the mean-field calculation, starting in the dispersed crystal phase. The critical particle densities $\rho_p^{\ast}$ determined by mean-field approximation and by  Monte Carlo differ slightly, but the discrepancy lies within the error of the Monte Carlo simulations. 

\section{Chaotropic Effect}
The addition of urea to an aqueous solution of hydrophobic molecules may affect the properties of the latter in a way that destabilizes aggregation of the non-polar solute~\cite{chitra}. In the case of protein solutions, this destabilization can result in a complete denaturation of the proteins, and even prevent their aggregation if the latter is due to hydrophobic interactions~\cite{kyte}. A high-concentration solution of urea is therefore often used as a protein denaturant.

The underlying cause of this process, known as the chaotropic effect, is generally believed to be a decrease in the order of the water structure ('chao-trope' = disorder maker), thus indirectly increasing the solubility of non-polar solutes~\cite{walrafen}. However, different attempts to discover how chaotropic agents perform this function have not yet been able to explain in a satisfactory manner the exact mechanism for disruption of the hydrogen bonds which stabilize the aggregate. In the remainder of this section, we adapt the MLG framework by including chaotropic cosolvent molecules with the aim of demonstrating the chaotropic effect in our model system.

Chaotropic substances are in general those which are less strongly polar than water. In aqueous solutions of non-polar species they act to reduce the number of possible intact hydrogen bonds between water molecules, both in the solvation shell and in the bulk, compared to a pure water-solute mixture. Within the adapted MLG framework, a straightforward approximation to the effect of a chaotropic cosolvent is to consider that its addition to strongly hydrogen-bonded, ``ordered'' clusters creates ``disordered'' clusters with additional broken hydrogen bonds and higher net energy. The creation of disordered states from ordered ones in the presence of a chaotropic cosolvent increases the degeneracy of the former at the expense of the latter.

Because the coarse-grained model treats each water site as containing a number of water molecules, the cosolvent molecules, which are generally rather small in comparison with the hydrophobic particles, may be included implicitly at each water site by adapting only the degeneracies of the energy levels of the water. The states of water clusters containing cosolvent are thus assigned the degeneracies $q_{os, u} =  q_{os} - \eta_s$ and $q_{ds, u} = q_{ds} + \eta_s$, and $q_{ob, u} =  q_{ob} - \eta_b$ and $q_{db, u}= q_{db} + \eta_b$, where $u$ denotes urea, which we adopt as an illustrative example of a small chaotropic cosolvent. The cosolvent is taken to affect only the number of hydrogen bonds formed, and not their strength, so in the bimodal approximation the energies of the states remain unchanged.  

The effect of a chaotropic cosolvent is much stronger in the bulk than in the shell. While in ordered, bulk water both tangentially and radially oriented hydrogen bonds may be broken, in shell water there exist no radially oriented bonds, and the ordered bulk configurations with radially oriented water have already become disordered configurations on substitution of the central water by a non-polar solute particle. The number of configurations available to be transformed from ordered to disordered by the addition of cosolvent is therefore much higher in the bulk than in the shell. Indeed, the number of configurations of ordered bulk states may be reduced almost to that of ordered shell water in the presence of a strongly chaotropic cosolvent, because the primary difference between the two is the absence of radial hydrogen bonds to the central molecule in the shell of a non-polar solute particle. For the calculations to follow we have chosen the values $\eta_b = 9.0$ and $\eta_s = 0.1$, which are found to be suitably representative of a water/solute/cosolvent system.

Within this approach the addition of a concentration $c$ of cosolvent to the water leads to an additional term in the partition function
\begin{eqnarray}
\nonumber
Z^u &= \sum_{\{n_i\}} &\prod_i(q_{os} e^{-\beta E_{os}} + q_{ds}e^{-\beta E_{ds}})^{n_i (1-\lambda_i) (1-c)}\\
\nonumber
&&\times (q_{os,u} e^{-\beta E_{os}} + q_{ds,u}e^{-\beta E_{ds}})^{n_i (1-\lambda_i) c}\\
\nonumber
&&\times (q_{ob}e^{-\beta E_{ob}}+q_{db}e^{-\beta E_{db}})^{n_i \lambda_i (1-c)}\\ 
&&\times (q_{ob,u}e^{-\beta E_{ob}}+q_{db,u}e^{-\beta E_{db}})^{n_i \lambda_i c}.
\end{eqnarray}
Here the concentration of urea in the bulk is assumed to be identical to the concentration in the first solvation shell of a hydrophobic particle. On defining ${\cal B}_u =  -\frac{1}{\beta} \ln[q_{ob,u}e^{-\beta E_{ob}} + q_{db,u} e^{-\beta E_{db}}]$ and ${\cal S}_u = -\frac{1}{\beta} \ln[q_{os,u}e^{-\beta E_{os}} + q_{ds,u} e^{-\beta E_{ds}}]$, the effective Hamiltonian function becomes
\begin{equation} 
H_{\rm eff}^u[\{n_i\}] = \sum_{i=1}^N [ \mu_{\rm eff}^u n_i + J_{\rm eff}^u n_i \lambda_i],
\label{eq:HeffUrea}
\end{equation}
where the effective interaction $J_{\rm eff}^u$ and the effective chemical potential $\mu_{\rm eff}^u$ are given by
\begin{eqnarray} 
J_{\rm eff}^u &= &{\cal B}-{\cal S}+c({\cal B}_u-{\cal B}+{\cal S}-{\cal S}_u)\\
\mu_{\rm eff}^u &= &{\cal S}+\mu+c({\cal S}_u-{\cal S}).
\end{eqnarray}
Because we are concerned with the hydrophobic interaction between solute particles, we again express the Hamiltonian in terms of particles $m_i=1-n_i$,
\begin{equation} 
H_{\rm eff}^{p,u}[\{m_i\}] = K^u - \sum_{i=1}^N [ \mu_{\rm eff}^u m_i + J_{\rm eff}^u (m_i-1) \lambda_i],
\label{eq:HeffUreaM}
\end{equation}
where $K^u=N\mu_{\rm eff}$ and $\lambda_i = \prod_{\langle i,j \rangle} (1-m_j)$. The effective interaction between the hydrophobic particles thus depends on the concentration of chaotropic agent in the solution. Because ${\cal S}_u>{\cal S}$ and ${\cal B}_u>{\cal B}$, $\mu_{\rm eff}^u$ is larger than $\mu_{\rm eff}$ and therefore it is easier to bring the non-polar solute into solution in the presence of chaotropic substances. In addition, the relation $({\cal B}_u-{\cal B})>({\cal S}_u-{\cal S})$ results in an increase in $J_{\rm eff}^u$ when raising the concentration $c$. Hence, the chaotropic particles support the repulsive force between non-polar solute molecules. The coexistence region of the phase diagram consequently shrinks with increasing $c$, and for extremely high concentrations may even disappear entirely.

\begin{figure}[h!]
\includegraphics[width=8cm]{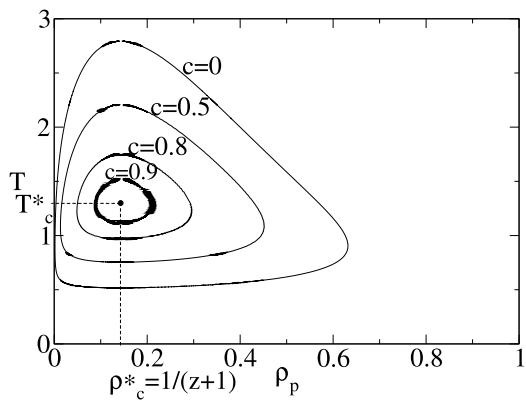}
\caption{$\rho$-$T$ phase diagram for a ternary system consisting of water, hydrophobic particles, and chaotropic cosolvent, obtained by mean-field calculation, showing coexistence curves for different cosolvent concentrations. The $\bullet$ represents a multi-critical point at the critical cosolvent concentration $c_c^{\ast}=0.935$ where the coexistence curve has shrunk to a single point.}
\label{FigBouleUrea}
\end{figure}

Fig.~\ref{FigBouleUrea} shows the effect on the coexistence curve for different urea concentrations $c$ using $\eta_b=9$ and $\eta_s=0.1$. The expected increase in solubility is confirmed, and the LCST and UCST approach each other with increasing cosolvent concentration. At a critical concentration $c_c^*=0.93$, the closed-loop curve shrinks to a single point ($T_c^*= 1.3,\ \mu_c^*=2.84$ and $\rho_c^*=1/(z+1)$), which represents a double critical point~\cite{rebelo}. For concentrations higher than $c_c^*$, the aggregation phase disappears completely.

\section{Discussion}
We have studied the aggregation of hydrophobic solutes in water using the bimodal MLG model, which we have adapted to describe a coarse-grained system where each site can be occupied by one or more molecules. One of the objectives was to reproduce the thermodynamic properties associated with the hydration and aggregation of non-polar solutes in water. For this purpose, Monte Carlo simulations were conducted to establish the phase diagram for the process, and the results were compared to a mean-field calculation.

As expected, both methods display clear phase transitions at a LCST and an UCST within a range of densities, and thus define a closed-loop coexistence region. Outside this region the system appears as a homogeneous particle-solvent mixture, while inside it a separation occurs into two phases of fixed (upper and lower) coexistence densities. The exact time evolution of the system after a quench into both the metastable and the spinodal regions at constant density has not yet been studied, but would offer interesting insight into liquid-liquid phase-ordering kinetics~\cite{bray, sara}.

At low temperatures ($T_0$ in Fig.~\ref{FigBouleMC}) the system is in the homogeneous region where the solubility of the solute is high and the solvent-induced effective interaction between solute  particles repulsive. On raising the temperature at constant density, the system shows a sharp transition to an aggregation state at the LCST $T_L$. In the equilibrium state at this temperature, clusters of aggregated, hydrophobic solute molecules with density $\rho_{c2}$ are suspended in nearly pure water (density $\rho_{c1} \simeq 0$). The aggregation allows the solute particles to minimize their exposed surface and to reduce the structural enhancement of the surrounding water, thus causing a positive entropy change. The effective interaction between non-polar particles becomes attractive above $T_L$, and therefore the formation of aggregates is preferred over the solution of single solute particles. Further heating of the system results in a second phase transition at the temperature $T_U$ where the hydrophobic particles disaggregate, and above this temperature one homogeneous phase is found. This process is dominated by the favorable entropy change of solvation in the whole system.

As can be seen from the $\mu$-$T$ phase diagram in Fig.~\ref{FigMuT}, the transition line for the Monte Carlo simulations and the mean-field calculation correspond rather well. They agree closely on the lower critical point, as well as on the densities above and below the transition temperatures. The upper critical point determined by Monte Carlo, however, lies at a lower temperature than the mean-field one. This result is not surprising, because it is well known that mean-field calculations, which neglect fluctuation effects, generally overestimate transition temperatures. The good agreement at low temperatures is rather a signature of the predominance of local effects which do not involve large fluctuations from site to site.

Apart from this difference, the coexistence curves from Monte Carlo and the mean-field calculations are qualitatively similar. At very low temperatures, however, the former show an additional crystal phase which cannot be explained by the latter, because it arises from spatial ordering of the hydrophobic solute and is therefore neglected by the present mean-field considerations, although this phase could be recovered within a more refined mean-field approximation. The appearance of this phase confirms the prediction of our analytical calculation at $T=0$.

\begin{figure}[h!]
\includegraphics[width=8cm]{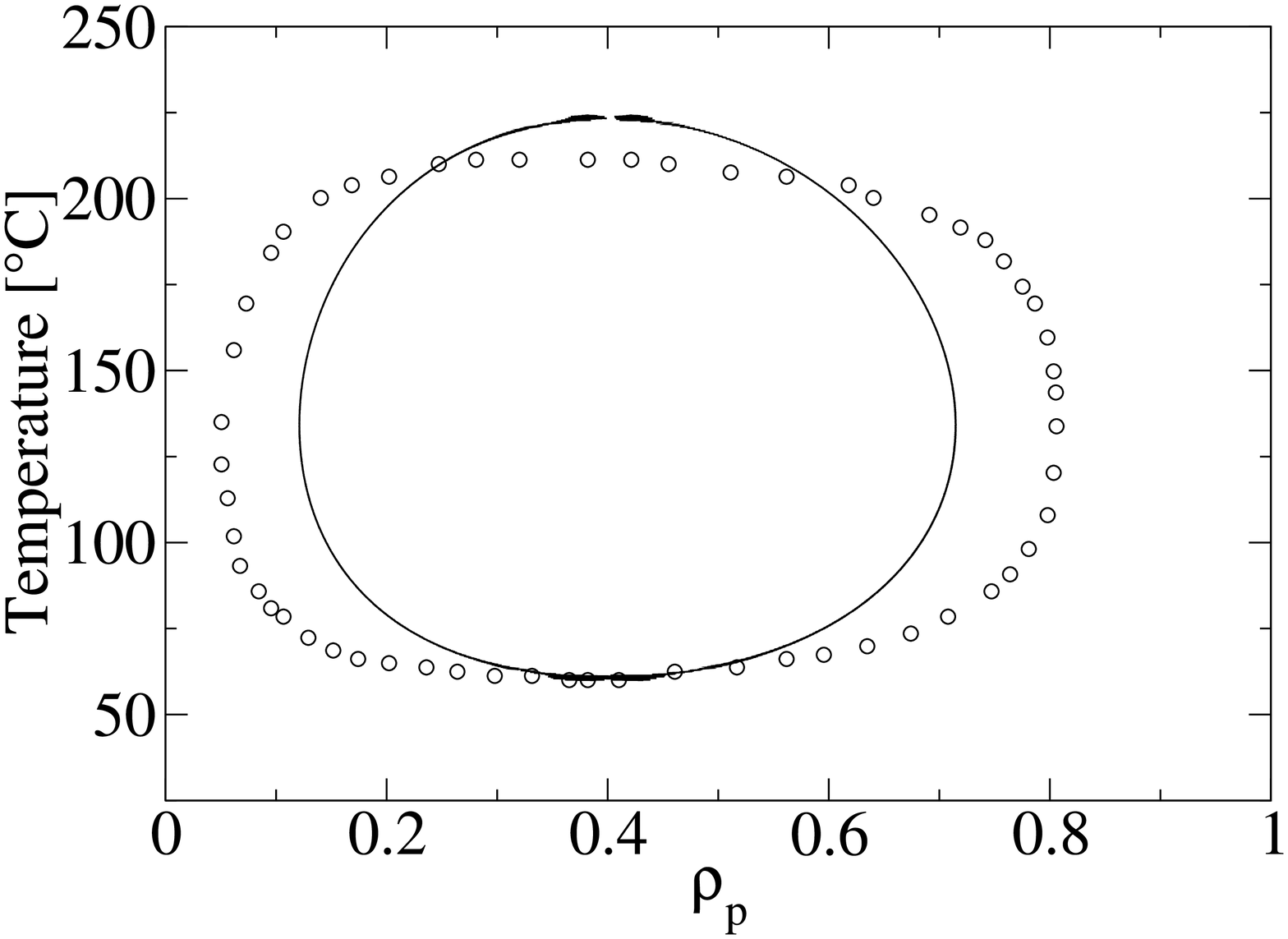}
\caption{Closed-loop solubility curve for nicotine/water, calculated within the mean-field framework by using the experimental value of the critical density and renormalizing by comparison with Monte Carlo simulations (see text). The curve shows good quantitative agreement with the experimentally determined solubility curve ($\circ$) reproduced from Ref.~\cite{davies}.}
\label{FigNicot}
\end{figure}

Encouraged by the qualitative and quantitative agreement of the mean-field calculation and Monte Carlo simulations, we have taken experimental values for the critical density $\rho_p^{\ast}$ in different systems to adapt the parameters of our model. Analytically, we have obtained the relation $\rho_p^{\ast}=\frac{1}{1+z}$ by mean-field calculations, and have confirmed this numerically by Monte Carlo simulations. Thus from an experimental value for $\rho_p^{\ast}$ we may extract an effective coordination number $z^{\rm eff}$, which is then introduced in the calculations. The critical density for the system nicotine/water is $\rho_p^{\ast}=0.4$~\cite{davies}, which results in $z_{n/w}^{\rm eff}=1.5$. This coordination number $z^{\rm eff}$ may be interpreted as the average number of hydrophobic solute molecules surrounding any chosen solute particle, which is relevant for the net effective hydrophobic interaction leading to attraction or repulsion. After changing the values for the energy levels to $E_{ds}= 3.4,\ E_{db}= 3.3,\ E_{ob}= -3.3$, and $E_{os}=-4.1$, the closed-loop coexistence curve is in good agreement with the experimental curve (Fig.~\ref{FigNicot}). The UCST is higher than the measured one, as expected because of  the mean-field nature of the calculation. The ratio between the mean-field result $T_{U,{\rm mf}}$ and the experimental value $T_{U, {\rm exp}}$ is $T_{U,{\rm mf}}/T_{U, {\rm exp}}= 1.25$. In fact this value agrees quantitatively with the ratio $T_{U,{\rm mf}}/T_{U,{\rm MC}} = 1.35$ which we obtain by comparing the mean-field calculation with the more accurate Monte Carlo simulations when both are performed for $z=6$ ({\it cf.} Fig.~\ref{FigMuT}). Using this as an effective scaling factor to renormalize the mean-field results for different $z^{\rm eff}$ yields good agreement with the experimental results for the nicotine/water system (Fig.~\ref{FigNicot}). We have repeated this procedure for the system poly(ethylene glycol) in water, which has a critical density of $\rho_p^{\ast}=0.15$ and a much larger molecular weight $M_w=3350$~\cite{bae}, and which resulted in a similar agreement with the experimental curve. 

In addition to studying hydrophobic solutes in pure water, we have analyzed the effect of a chaotropic substance in the solution. The destabilizing effect can be attributed indirectly to a disordering of water molecules by chaotropic cosolvents associated with a rearrangement of the energy states of the water. As a first approximation, we have included the chaotropic agent in liquid water by increasing the number of disordered water states with respect to ordered ones. A mean-field calculation for the model with cosolvent shows the expected increase in solubility of the hydrophobic particles. At concentrations higher than a critical value, the aggregation phase disappears completely. Monte Carlo simulations confirm this behavior. 

Experimentally, high concentrations of chaotropic substances are required to cause destabilization of aggregates of hydrophobic particles. As an example, the solubility of the highly hydrophobic amino-acid phenylalanin at room temperature is doubled in an $8$-molar urea solution, which consists of an equivalent volume fraction of urea and water. While within the current framework it is difficult to express the cosolvent concentration explicitly because the model contains more than one water molecule per site, the results are qualitatively correct. A cosolvent concentration of at least $50\%$ is required to reduce significantly the extent of the solubility region (Fig.~\ref{FigBouleUrea}). 

We should emphasize that in the mean-field analysis presented here the concentrations of chaotropic agents in the bulk and in the first solvation shell were assumed to be equal. However, chaotropic molecules are found preferentially in the solvation shell of non-polar solute particles. Considering two different concentrations may be expected to provide better insight into the exact mechanism of the chaotropic effect~\cite{moelbert}.

Overall, we have shown that qualitative features of the liquid-liquid demixing process of hydrophobic aggregation, as well as of the chaotropic effect, may be explained satisfactorily within a simple model for aqueous solutions of non-polar particles by including hydrophobic interactions only in terms of changes in water structure. Although the explicit terms of model describe solely the states of water molecules in solution, we have demonstrated that it contains implicitly both two- and even many-particle interactions between hydrophobic solute molecules. The complete density-temperature phase diagram was established, by both analytical and numerical techniques, and illustrates the characteristic properties of hydrophobic aggregation.

\acknowledgments
We are grateful to B. Normand for helpful discussions and comments. We wish to thank the Swiss National Science Foundation for financial support through grant FNRS 21-61397.00.

\end{document}